\documentclass[conference]{IEEEtran}
\IEEEoverridecommandlockouts

\usepackage{cite}
\usepackage{hyperref}
\usepackage{todonotes}
\usepackage{soul}
\usepackage{color}
\usepackage{xspace}
\usepackage{balance}

\newcommand{\hassanbox}[1]{%
  \smallskip%
  \noindent\framebox{%
    \begin{minipage}[c]{0.97\linewidth}%
      \begin{center}%
        #1\par%
      \end{center}%
    \end{minipage}%
  }%
  \smallskip%
}

\begin{document}
\title{Sourcerer's Apprentice and the study of code snippet migration}
\author{\IEEEauthorblockN{Stephen Romansky, Abram Hindle}
  \IEEEauthorblockA{Department of Computing Science, University of Alberta\\
    Edmonton, Canada\\
    {romansky, hindle1}@ualberta.ca}\\
  \and
  \IEEEauthorblockN{Cheng Chen, Baljeet Malhotra}
  \IEEEauthorblockA{BlackDuck\\
    Burnaby, Canada\\
    {cchen, bmalhotra}@blackducksoftware.ca
    }
}

\maketitle
\footnotetext[1]{We would like to thank Vaibhav Saini for help to get SCC running.}

\footnotetext[2]{Get our code and data at the following URL: \url{https://github.com/SRomansky/SourcererCC.git} (TODO curate and publish everything for camera ready)}

\begin{abstract}
  On the worldwide web, not only are webpages connected but source code is too.
  Software development is becoming more accessible to everyone and the
  licensing for software remains complicated. We need to know if software licenses
  are being maintained properly throughout their reuse and evolution.
  This motivated the development of the Sourcerer's Apprentice, a
  webservice that helps track clone relicensing, because software
  typically employ software licenses to describe how their software may
  be used and adapted.
  But most developers do not have the legal expertise to sort out license conflicts.
  In this paper we put the Apprentice to work on empirical studies that demonstrate there is much sharing between StackOverflow code and Python modules and Python documentation that violates the licensing of the original Python modules and documentation:
  software snippets shared through StackOverflow are often being relicensed improperly to CC-BY-SA 3.0 without maintaining the appropriate attribution.
  We show that many snippets on StackOverflow are inappropriately relicensed by StackOverflow users, jeopardizing the status of the software built by companies and developers who reuse StackOverflow snippets.
  \textsuperscript{1,2}
\end{abstract}

\section{Introduction}

When software is written it is covered by copyright granting the
author exclusive rights to the distribution of their software.
Software typically must be licensed to other parties for it to be used,
distributed, and sold. Software can be licensed by
developers to impose or alleviate restrictions on how it may be
reused. Open-source software licenses typically seek to enable the
free reuse and distribution of software provided that attribution to
the authors is given.
Code reuse results in numerous ``code clones": exact or near-exact code snippets or files occurring within multiple software projects that are still licensed.
Large programming sites
like StackOverflow shares source code in answers and questions~\cite{SO}
as well as sites like GitHub that share code in publicly hosted software
repositories~\cite{github}.
Code on StackOverflow is typically claimed to be opensource by StackOverflow's
terms of service, but the code might have come from elsewhere and
someone else. We raise the question, ``Can we trust the license of
code shared on StackOverflow to be accurate?''

A common open source license violation is the lack of attribution, most Free/Libre Open Source Software (F/LOSS)  licenses require that the authors who wrote the code are attributed in documentation, in the source code, or in startup messages. Not attributing the opensource copyright holder violates the opensource license.
Thus using the wrong license or misattributing code can be costly because:
\begin{itemize}
\item a developer or company can lose the rights to use, reuse, and distribute source code and software they rely on;
\item a developer or company a developer may be required to distribute their proprietary source code unexpectedly, if a copy-left license was included;
\item or, the developer or company may be sued for copyright infringement~\cite{bltj, itworld, scompliance, lexology}.  
\end{itemize}

We investigate the code clones created between StackOverflow and Python modules
as well as StackOverflow and Python documentation to determine if developers
are copying common reference material without proper relicensing or attribution onto StackOverflow.
We find the relationship between the source of code clones like Open Source Software and StackOverflow is bi-directional, rather than uni-directional~\cite{yang2017stack}, and fraught with license inconsistencies.
For instance, copying code to ask a question or share an answer on StackOverflow relicenses the shared code to the \texttt{CC-BY-SA 3.0} license~\cite{SOisCCBYSA3, ccbysa3}.
For instance, \texttt{GPLv3} code cannot be posted
on StackOverflow due to incompatible relicensing.
It is important to understand how developers interact with community-driven tools like StackOverflow where 66\% of software developers who use the StackOverflow service are unaware of the license it imposes on their posted code~\cite{baltes2018usage}.
In this paper we highlight a severe problem that code posted to StackOverflow often has incompatible licenses, but also that license terms are breached by the lack of attribution---this imperils the reuse of StackOverflow code as end-user developers could be liable for copyright infringement.

Our work also discusses our extension of SourcererCC to create the Sourcerer's Apprentice: that detects code clones with possible relicensing issues.
The Sourcerer's Apprentice
can be used by any developer interested in checking if their code base has copied artifacts with candidate license-inconsistencies from open source
repositories. We demonstrate this web service by: detecting if students have submitted homework
 solutions plagiarized from StackOverflow; to detect if students have copied home work from each other; and to perform clone detection on non-source based repositories of code blocks like StackOverflow and software documentation (Python 2.7) to find the relicensing conflicts from copying reference material.
With the help of the Sourcerer's Apprentice our contributions are:
\begin{itemize}
\item We show relicensing conflicts between Python Modules and Python Documentation and StackOverflow;
\item We show that the flow of code between StackOverflow and Open Source Software is bi-directional;
\item Implementation of a web service for clone detection and detailed reproducibility performance improvements to existing tools;
\end{itemize}

\section{Background}
\label{background}

We cover preliminary material in this section that introduces code
clones, software license detection, and subsets of relevant web
protocols used to build our web service. We also provide short
descriptions of state-of-the-art code detection tools and license
detection tools such as SourcererCC~\cite{sajnani2016sourcerercc} and Ninka~\cite{german2010sentence} that are used in our
work.

\textbf{Software licenses:}
are applied by developers to constrain and enable future reuse.
By default the software creator is the copyright owner.
As exclusive copyright holders, developers use licenses to distribute software
to clients under various terms.
Developer license their work with licenses from two popular categories
of licenses: F/LOSS and proprietary. Open source focuses on having
shareable and editable code, while proprietary licensing restricts who
may access the code, run the binaries, and whether users may alter it.
Researchers have also investigated when, and how, license changes occur
in software projects through revision history and issue
trackers~\cite{vendome2017license}.

\textbf{License conflicts:}
License conflicts occur when two or more licenses impose restrictions on
each other that cannot be met. 
This can happen between any two licenses, regardless
of their categories, based on their requirements. 
For instance, the GPLv2+
license is compatible with the GPLv3 but GPLv3 is not compatible with
GPLv2~\cite{gpllicenses}.  
Licenses are hard to interpret and often
require legal expertise. 
Therefore, work has been invested in
software license analysis, understanding license evolution, and conflict detection~\cite{german2010sentence, german2012method, german2009license, van2014tracing, mlouki2016detection, vendome2017machine, almeida2017software}.
We follow suit with our investigation of
code reuse and improve existing automated license detection.

\textbf{Code clones:}
Code clones occur when two code segments, of 1 or more code tokens, appear to
be
similar to one another under a given comparison function. Similarity could mean
syntactically, that the two fragments appear to have a similar number of 
features, corresponding variables, or that there are matching sequences of
code tokens in each fragment. 
It could also mean semantically, where two
fragments of code look dissimilar to one another, but each one performs similar
operations when it is run. 
It is up to the investigator to determine what
the similarity function is for relating two code fragments.

Conventional, and accepted, literature defines $4$ types of code clones:
\textbf{Type-1} are identical code segments ignoring white space
and comments; \textbf{Type-2} are type-1 clones with the additional exceptions
that the segments can have modified identifiers, literals, and types from one
another; similarly, \textbf{Type-3} clones are type-2 clones with the
additional exceptions that the segments can have added or removed lines; while,
\textbf{Type-4} clones are code segments that perform the same actions,
but are syntactically different~\cite{roy2007survey, german2010sentence}.
In prior work, code segments are often referred to as code blocks and are
extracted from whole functions or whole
files; but, we are interested in the license of any segment of
code, even if these are incomplete
software components~\cite{roy2007survey, sajnani2016sourcerercc}.

Code similarity, or code clones, play an active role in the development life
cycle.
Kasper \emph{et al.} shows some clones cause more technical debt
than they resolve, while others simplify prototyping in feature
development~\cite{kapser2006cloning,kapser2008cloning}.

It is also possible to use code clones to study projects that we do not
own~\cite{german2009code}.
We can view how much code is duplicate, or what is frequently reproduced in
software projects, and we can try to help developers by making generalizations
based on our observations to create new conventions or functions based on our
analysis.

\textbf{Software communities and code sharing:}
It is possible to study software communities to understand code clone growth and evolution, to see where license conflicts can be created, and to view
clones and licenses together.
Developers can write alone, or with others. The internet makes it easier for
developers to work together and to share code with one another. However, code
sharing requires licensing. GitHub provides developers with a place to share
their projects under the constraint of the terms of service~\cite{githubtos},
but developers are free to license their software
projects however desired. StackOverflow provides developers with a place to
share their development questions under the constraint that any posted code
will be licensed under the \texttt{CC-BY-SA 3.0} license~\cite{SOisCCBYSA3}.
GitHub and StackOverflow can help developers further their project goals with
social interactions.

Another service that developers have used in the past, and still use,
to share questions and projects were mailing lists.
Mailing lists are not required to have a user agreement or terms of use;
however, some show disclaimers about usage and posted
content~\cite{debiandisclaimer}. Developers have also used
websites to share code in the past like \texttt{freshmeat.net}.

\textbf{License detection -- Ninka:}
Ninka is a license detection tool that uses regular expressions to detect
112 different licenses~\cite{german2010sentence}. It works by attempting to
extract comments that occur near the start of a file. This is the position a
license is typically placed in a source file. Then, regular expressions are
applied to this comment block to try and match given licenses. If success
occurs and a license is matched then it gets reported by the program. However,
failure can happen in two ways: a license is found but unknown, or no license
is found. In these respective cases the program will report UNKNOWN or NONE for
the license of the given file.

\textbf{Representational state transfer (REST):}
REST is a loose communication protocol, used by web services~\cite{REST},
that tries to represent state and actions using HTTP concepts such as HTTP verbs (\texttt{GET}, \texttt{PUT}, \texttt{POST}) and treats entities as HTTP Responses and URIs as names.
Our system is built up on REST enabling composition of automation tools and UIs.

\textbf{Clone detection:}
Many authors have created code clone detection tools~\cite{kamiya2002ccfinder, jiang2007deckard, gode2009incremental, cordy2011nicad, wang2012can, lin2015clone, toomim2004managing}, but
we use a tool called SourcererCC (SCC) to perform code clone detection in our
work.
SCC was advertised as an ``internet scale'' code clone detector having
superior performance to its competitors.
Researchers have also worked on studying and generalizing code clones once detected~\cite{kim2009discovering, duala2007tracking, lin2014clonepedia, kapser2004aiding, inoue2012does}.
Prior work
has shown that SourcererCC performs well on big
code~\cite{sajnani2016sourcerercc}; therefore, in this study we omit comparison against
distributed code clone finders like D-CCFinder~\cite{livieri2007very}. 
D-CCFinder wrapped CCFinder in a distributed manner across a cluster.

SourcererCC works in 3 stages: (1) parse code into a format interpretable by the tool;
(2) create an index from the parsed source code to speed up clone detection; and (3)
perform clone detection~\cite{sajnani2016sourcerercc}.
In stage (1) the parser is not specifically part of SCC it abides by the
format used in it though. Stage (1) uses a parser to tokenize files into
blocks interpret-table by the clone detection algorithm. Researchers can use 
any convention to define a code block that is needed. Prior work defines code
blocks from file level~\cite{sajnani2016sourcerercc} and function
level~\cite{yang2017stack}. Parsed blocks record the unique tokens from the
code and the number of tokens in each segment.
Stage (2) creates an index that matches code tokens to blocks that contain
the tokens. This allows the clone detection process to quickly find blocks
that are related by looking up blocks that contain similar tokens.
Stage (3) compares blocks from a query set to a corpus, a data set, to determine if there
are clones between the two sets of code blocks. This is done by
checking if each block has an overlap of 80\% of it's tokens with any block
from the other set. We picked 80\% for this work, the overlap can be
configured by the user. The result of this process is a list of clones that
researchers may interpret.

Using SourcererCC at function-level granularity, Yang \emph{et al.} provide a
comprehensive study of the Python clones detected between GitHub and
StackOverflow~\cite{yang2017stack}, which shows high reusability of Python
code between StackOverflow and GitHub.
Code is being reused from StackOverflow regardless of the license.
Given this prior qualitative study on StackOverflow and GitHub, we focus on the licenses associated with migrating code clones as well as extend Yang \emph{et al.}'s work by using a \emph{pip} based data set to replace the GitHub data set.

\section{Related work}
\label{related work}

This section discusses recent applications/tools
 in the field of license detection, and license compliance.

Prior work has been accomplished on the task of detecting license inconsistencies between software projects by Wu \emph{et al.}~\cite{wu2015method, wu2017analysis}. Wu's method involves looking at
the license comment included in each source file and their mutation over time.
Wu \emph{et al.} use file-level clone detection to track the evolution of
a file with respect to code changes and Ninka to track license
evolution~\cite{wu2017analysis}. Wu's approach is extended and applied to
analyze license inconsistencies in large-scale F/LOSS
projects~\cite{wu2017developers}. We extend the application of
Ninka for license inconsistency detection by checking the package-level files
that developers use to license every file in the project in addition to
checking the license of the individual files.

German \emph{et al.} have also studied code clones and their
licenses to uncover how copied-licensed code evolves in parallel with respect
to it's point of origin~\cite{german2009code}.
This is highly relevant to our
work; however, we are interested, specifically, in the clone creation period at
this point in our tools development.
As well, Davies \emph{et al.} propose and demonstrate a method for identifying the
origin of software components, like libraries, to make it easier for developers
to track and verify where their software dependencies come
from~\cite{davies2011software}. 
We would like to answer whether or not a clone was created that violated the
license of the originating code snippet in the clone pair.

\section{Methodology}
\label{methodology}
In this section, we describe the overall architecture of our service;
the research and empirical questions we would like to investigate with
the tool; and we show how the tool produces easy to read reports of
software clones and licenses.
We start
by introducing several of the code clone identification tasks (in
Section~\ref{problems}) that we would like our service to perform. We provide a
detailed account (in Section~\ref{webservice}) regarding our process for
extracting software licenses and code clones from input source code.
Section~\ref{clone-license} describes our approach to aggregating the
code-clone and licensing information into an easy-to-read HTML report.

\subsection{Problems with copy and pasted code}
\label{problems}
\newcommand{\caseA}{Case 1[Student--SO]\xspace}
\newcommand{\caseB}{Case 2[Student--Student]\xspace}
\newcommand{\caseC}{Case 3[Pip--SO]\xspace}
\newcommand{\caseD}{Case 4[Docs--SO]\xspace}
In software development it is not uncommon to have duplicated code. There are
many reasons to make duplicate code that we will not be focusing on. Instead,
we are interested in identifying when copy-pastes happen and the licensing
complications these simple actions incur. Specifically, we are interested in the following use cases and scenarios:
\emph{\caseA}  copy-paste detection between student assignments and StackOverflow posts;
\emph{\caseB} copy-paste detection between student assignments; 
\emph{\caseC} copy-paste detection with license compliance between code submissions and a static corpus of source code;
and
\emph{\caseD} copy-paste detection with license compliance between documentation and StackOverflow.

It is important that students attribute their sources when completing
their work, as in \emph{\caseA}, to avoid plagiarism and license
violations. Such a habit may carry-forward into professional software
development where license violations can be costly. The popularity of
software development courses has also made it more cumbersome to
detect plagiarized assignments, therefore \emph{\caseB} would ease the burden
placed upon teaching staff when grading homework and aide in upholding
academic integrity. In \emph{\caseC} we wish to accomplish license
violation detection in general code corpora. This would help anyone
with a software project check if any of their code has been
copy-pasted from public software repositories of code.
With \emph{\caseD} we want to check if unclear code migration occurs between
StackOverflow and documentation because these two information sources use
different, incompatible, software licenses and would portray license compliance
violations if unclear migration exists.

\subsection{Web service}
\label{webservice}

\begin{figure}
  \includegraphics[width=0.5\textwidth]{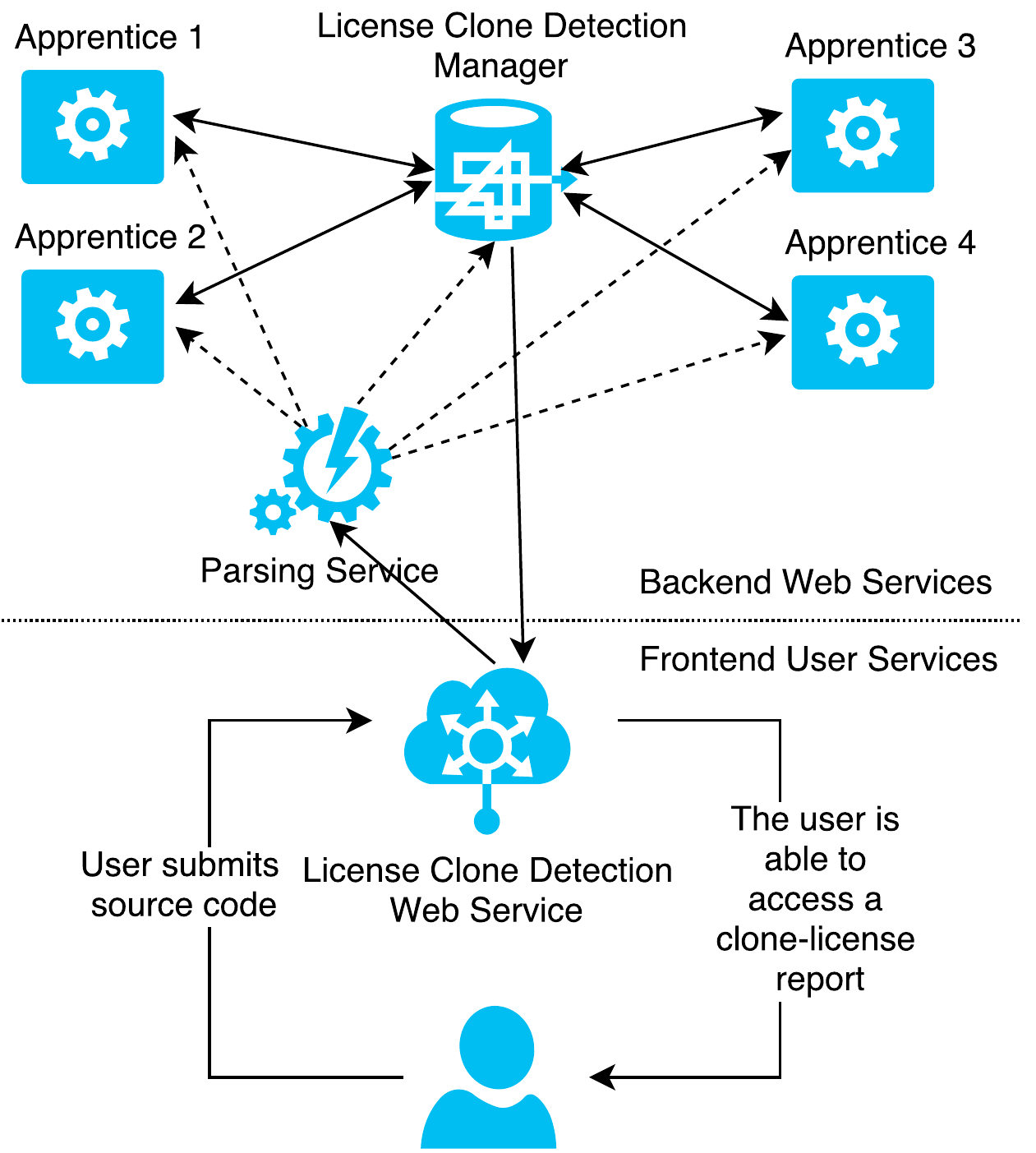}
  \caption{Communication diagram of license-clone web service containing:
    $1$ parser web service; $1$ license-clone management web service; and
    $4$ clone detection Sourcerer apprentice services. The arrows represent
    flow of information throughout the microservices. This includes the parser
    sending parsed source code to the apprentices and their manager, as well as sending reports
  to the service-user.}
  \label{fig:flow}
\end{figure}

\begin{figure*}
  \centering
  \includegraphics[width=.7\textwidth]{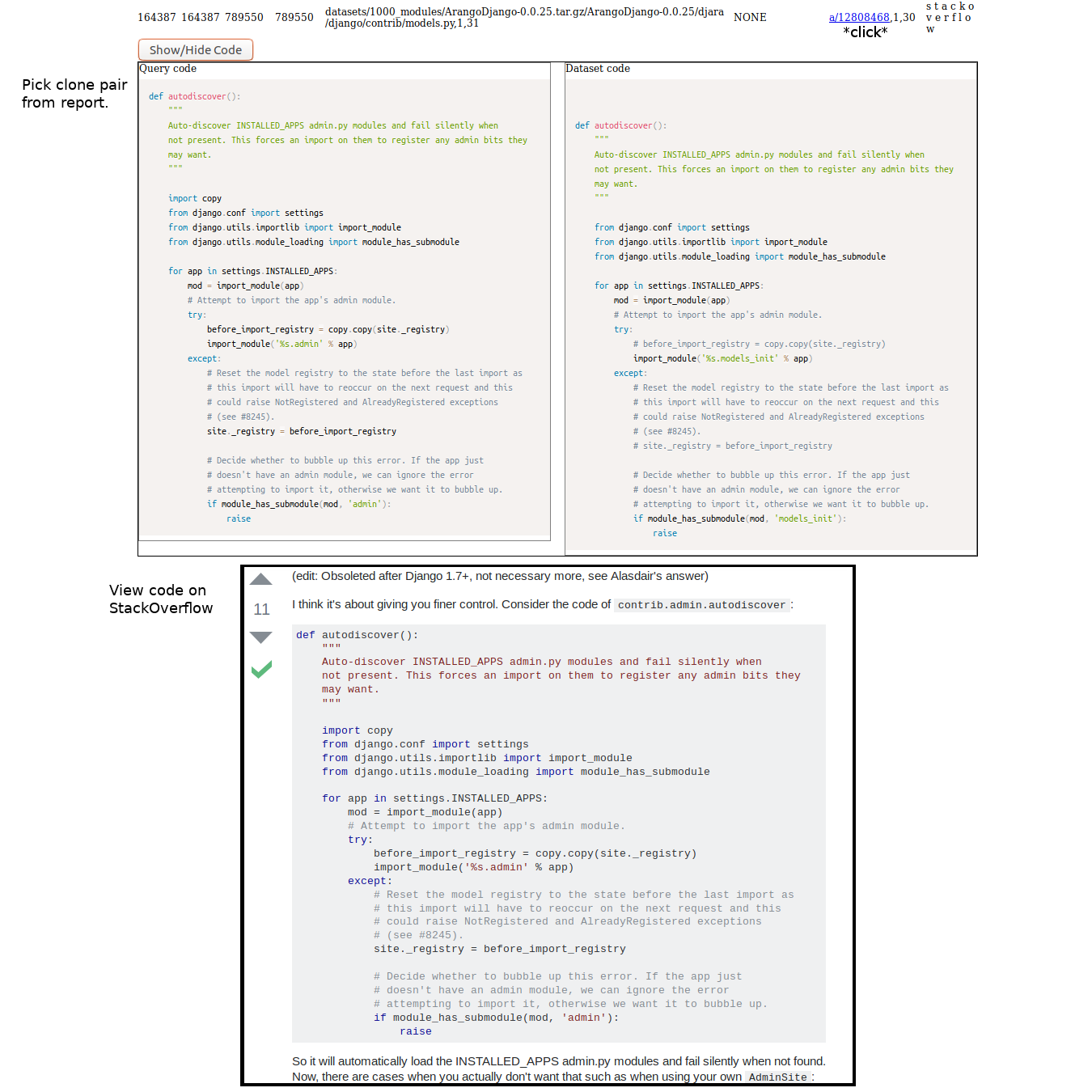}
  \caption{A user looking up sources from a detected clone pair of candidate
    copy-paste. The user begins by wanting to view the context of a
    StackOverflow post from the report. The user clicks the URL in the report
    and this opens the web service to view the StackOverflow post.}
  \label{fig:usage}
\end{figure*}

We made a tool that takes source code, submitted by a user, and
finds clones in it with respect to a known corpus of source code that is labelled with their respective software licenses. Figure~\ref{fig:flow} shows this
process. Figure~\ref{fig:usage} shows an example code clone report and how to access the code snippets place of origin.
From Figure~\ref{fig:flow} it is shown that we actually have several
web services that interact together. The rest of this
subsection describes these web services.

The parser service, of Figure~\ref{fig:flow}, accepts code from a user and
parses it into the format required by SourcererCC. Once the parser has
processed the source code it sends it to apprentices or the management service.
The parsed source code can be compared against a larger corpus to find duplicates between
itself and the larger corpus. Or, the parsed code can act as the corpus and
have user queries compared against it.
If the parsed code is sent to the apprentices, it is used as a corpus, which users
can
query for clones. If the parsed code is sent to the manager, it can be used as a query
set, which users can compare to the corpora held on each of the separate
apprentices.
Apprentices typically host independent corpora or independent subsets of the main corpus.
The parser applies license detection, using Ninka, in addition to the parsing
and formatting of source code for SourcererCC. This provides the other
web services with licensing details on the code blocks being used for clone
detection.

The management service, of Figure~\ref{fig:flow}, maintains a list of query
sets that can be sent to the apprentices.  These are sets of parsed source code which are sent
to all of the clients to distribute the workload of clone detection, compared
to having only 1 apprentice performing all of the clone detection work on 1
machine. The apprentices can be running on different devices on different
networks.
The management service also maintains a list of active apprentices.

The manager also maintains query results, a list of the clone-license results. When an
apprentice
has completed a query against a corpus it sends detected clones and their
licensing information back to the manager. The manager then generates a report
which shows the code, where the code is from, and the licenses of each code
block (from the query set and the corpus of the apprentices.) The manager
then
displays these reports to the user by sending them to the users browser.
This allows a user to view clones, the source code of the clones, and the
licenses of the detected clones. Part of an example report is shown in
Figure~\ref{fig:usage}. Queries can range in size from containing 0 to several
million code snippets.

The apprentices, of Figure~\ref{fig:flow}, run SourcererCC. Each apprentice
loads a local corpus, distributed from the parser, and
compares queries to the local corpus when the manager requests.
The apprentices allow us to
distribute our  workload, horizontally, across networks.

\subsection{Clone-License Identification and Reporting Detailed}
\label{clone-license}

This subsection details the assembly of the clone detection apprentices and the
parser. We heavily build upon others work like SourcererCC~\cite{sajnani2016sourcerercc} and Ninka~\cite{german2010sentence}.

To assemble our parser we used a Python library called \texttt{ASTTokens}~\cite{asttokens} which allowed us to tokenize Python code into the required format and it allowed us to extract the source code for our report generation.
The code blocks are considered to be a
specific context of code. In our application we parse code blocks that are
file-level, module-level, and function-level:
code contained only in a given function.
The SourcererCC tool is then able to compare code blocks to detect syntactically similar clones~\cite{sajnani2016sourcerercc}.

While extracting the code blocks, the parser is also able to apply a
modified version of Ninka~\cite{german2010sentence} called
\emph{NotQuiteNinka}. Ninka is able to extract the license for each
function, module, or file that we are reformatting. We wrap
Ninka in \emph{NotQuiteNinka} such that it also checks the pip module
package files for license information, because not all of the Python
source files contain their respective licenses. An example of the primary
package structure
that \emph{NotQuiteNinka} takes advantage of are files called LICENSE which
specify the license of the whole module, unless otherwise specified, and are
plain text documents. The modification of \emph{Ninka}
enables us to approximate the licenses of the software clones we study. Our
evaluation provides details on the improvements added by checking
package configuration files.

In addition, when extracting the code blocks, the parser is able to extract
the time that the code block was last modified. Time retrieval can be
accomplished by checking the context the code block occurs in for information
such as: posted time; last time the file was modified; and checking for
revision control time data. Time allows developers to approximate which code
came first and is the original copy~\cite{davies2011software}.

\subsection{Collected Datasets: documentation, SO posts, pip modules}
\label{datasets}

With tools to extract the code clones and their licenses, we need a data set
to study software clone usage with licenses. To accomplish this we use the
Python package management tool, \emph{pip}~\cite{pip}, to collect a data set of open source
Python code. We also collect a data set of developer code snippets from the
StackOverflow~\cite{SO} web service. 
We are able to combine the SourcererCC tool with Ninka to check if there is
code shared between StackOverflow and pip packages. This is typically a license conflict as  
the StackOverflow code is licensed under 
\texttt{CC-BY-SA 3.0}~\cite{SOisCCBYSA3}, which
is incompatible with many F/LOSS copy-left licenses.
Our web service is able to provide a view of the code from the corresponding
StackOverflow post as well as a direct URL link to the StackOverflow post.
Developers can submit many different types of code snippets to the service such
as the examples detailed in Section~\ref{problems}.

From \emph{archive.org}~\cite{SOData} we collect a dump of all StackExchange boards which
contains StackOverflow as of December 1st, 2017.
The parser searches for module and method level code blocks inside the posts.

From GitHub~\cite{githubPython} we collect a copy of the Python 2.7 repository and its documentation.
The documentation is parsed for each module and method level code block.

From the PyPi~\cite{pypi} we collect a list of available pip packages.
We select $5000$ packages to download randomly and use the pip download command to collect them.
$5000$ was chosen arbitrarily, we partitioned them into data sets of size 10, 100, 1000, and 5000 modules that were named 10m, 100m, 1000m, and 5000m respectively.

The front-end web service provided by the Apprentices accepts compressed directories
and searches these for Python code from our parsers.

\section{Evaluation}
\label{evaluation}

To evaluate
our tool we provide a summary of our benchmark results and evaluate the 4 use
cases outlined in Section~\ref{problems} that are: \emph{\caseA}, evaluating a
student-esque project against StackOverflow to check for copied
content (in Section~\ref{student}); \emph{\caseB}, comparing
student-esque projects against student-esque projects to check for
plagiarism, like in a class room setting, in Section~\ref{students};
\emph{\caseC}, to query general software projects against other
software projects to check for software migration in terms of
copy-pastes and code clones (in Section~\ref{sample};)
and, \emph{\caseD}, to determine if clones exist between documentation
and StackOverflow with license compliance issues.

\subsection{Software modification benchmark summary}
With the original application it took $44.22$ hours to execute
a query of $5000$ against $5000$ Python modules. After our upgrades it took
$56.4$ seconds. These time measurements are averaged over $5$ runs.
To improve performance we did 3 things: 1) we wrote a benchmark test suite to
evaluate whether new modifications helped improve performance; 2) we cached
data that can be reused between separate queries on the same data set; and 3)
we found bottlenecks in the pipeline used to process clone candidate
evaluation. We used our benchmarks of multiple simulated clone detection
queries against varying number of modules to guess and check what part of the
software was causing a slow down. We found that indexing parts of the code
blocks took a while for each query and therefore cached this task by
daemonizing the clone detection tool. We also found that detecting if a code
block from the query set has any clones takes a while to go through the queue
of clone candidate evaluation stages; therefore, we made each code block into
it's own asynchronous job that is managed by the java runtime environment using
Java 8 Stream objects.
Java 8 Stream objects are JVM-built-in data
queues with multi-threading support for parallel queue entry
processing~\cite{streams}. This means users do not have to worry about
performance tuning the manually constructed task queues.

\subsection{\caseA:
 copy-paste detection between student assignments and StackOverflow posts}

\label{student}

Project Euler problems have been attempted by many developers and some of their
solutions appear on GitHub~\footnote{\url{https://projecteuler.net/}}, we use
30 projects from GitHub to emulate a set of well-defined student projects.
The projects were selected by querying ``\texttt{project euler solution}'' on GitHub and selecting the top $30$ Python repositories  ranked by stars.
These repositories were cloned and parsed with our Python parser for clone
detection.
One of the top $30$ projects was not a programming exercise solution set, so
it was ignored.
We used the Apprentice to check if any of the $29$ student projects contains code copy-pasted from StackOverflow.

\textbf{The result:} we found that it was possible to check if homework had
been copied from StackOverflow; however, we did not find any of our exercises
copied from StackOverflow.


From the $29$ projects we extracted $7,107$ Python code blocks.
We queried the StackOverflow code blocks with the Python project code blocks using the Apprentices. 
We did not find any code clones between the project code blocks
and StackOverflow.
Therefore, it can be assumed that the students developed
their own solutions to the Project Euler problems without copying code
from StackOverflow.
The lack of
clones may be caused by the brevity of the programming exercise solutions
and the clone detector requiring a minimum number of tokens in each code
block.
However, the number of clones and the amount of noise/non-meaningful clones
grows when the minimum number of tokens per code block is reduced.
Another possible scenario that could have occurred is that solutions were
discussed line-by-line instead of as a single-whole code entry in the StackOverflow posts.
This use case demonstrates that it is
possible for web service users to check if their code conflicts with existing
web content like that of StackOverflow.

\subsection{\caseB: copy-paste detection between student assignments}

\label{students}

Using the Apprentice and manual inspection of code clones, we check if code was copied between the 29 student projects used in \emph{\caseA}---Project Euler solutions.


\textbf{The result:} the tool is capable of detecting duplicate
homework and candidate duplicate homework.

To perform this case study, we found $7,107$ code
blocks total from the $29$ exercise sets of \emph{\caseA}. We performed an intra-data set query on the 29 projects with themselves using the Apprentices and found $45,450$ clones. We did not find forks of projects in the data set. 
One repository does not share clones with any other of the 28 projects, but
this repository has $44,942$ clones within itself, therefore we ignore it in
our analysis for copy-pasted code between repositories.
The remaining 28 projects had $508$ clones.

We manually inspected the $508$ clones: there are many intra-project clones showing the
students copied their own code.
There were still inter-project (between project) clones found,
in one example we saw $52$ of the copy-pasted clones occurred
between the $2$ program exercise sets; one of the students appears to
have copied many solutions from another student with minor modification.
From the inter-project clones, we also found $32$ similar solutions in the clone report that were semantically
similar yet syntactically distinct---which makes sense given that they are solutions to  the same exercises.
We found $16$ copy-pastes that showed duplicate code between
programming exercises and we found $1$ detected clone that had unrelated code
blocks in it, between two distinct repositories.

We did not investigate license conflicts or attribution in this case study.
One weakness of the Apprentice tool is that it does not check for attribution.
This study demonstrates that the Apprentice tool can detect
plagiarized homework assignments.

\subsection{\caseC: copy-paste detection with license compliance between code submissions and a static corpus of source code}

\label{sample}


We want to know if code gets copy pasted between Python modules and StackOverflow
(clone migration), because these two code contexts can use
differing-incompatible licenses.
To study this question, we extract Python code from pip modules and
from each Python post on StackOverflow.
Each post containing code is treated as a separate file and therefore generates at least 1 code block.
We sampled code clone pairs based on sizes to answer our question. 
We manually investigated each clone pair to determine if we felt they looked like they were copy-pasted.
Manually, classification was done by reading the code pairs and labelling them as type-1, type-2, or type-3 clones and whether
the code pairs were semantically similar if any modification had occurred. 
We categorized clones into three groups to begin: small (1-10 lines); medium (11-20 lines);
and large (greater than 20 lines).
Other work by Wu \emph{et al.} also supported the idea that it is hard to
determine who created a small-clone and who the rightful owner is in their
file-level-clone based work~\cite{wu2015method}.

\textbf{The result:} we find that 24\% (15/63) of medium sized (11-20 lines long) clones are copy-pasted;
6\% (4/63) of large (more than 20 lines) clones are copy-pasted; and that these copy pastes have conflicting software licenses with StackOverflow's license.

We focus on finding the rate of copy-paste occurrence in
the medium and large size clone groups. We calculated a sample size of
$63$ for the 
power calculation for two proportions (same sample sizes). This would
enable us to detect medium effect sizes ($0.5$), significance level of $0.05$ with a power of 80\% to have
sufficient representative power from the medium-sized and large-sized
clone groups. We investigate prior work for sampling techniques, to
study the properties and features of code clones; however, we did not
find any prior work that used sampling~\cite{wu2015method,
  wu2017analysis, yang2017stack}.

From a clone report of 1000m querying our StackOverflow data, using the Apprentices, we found via manually reading the clones that
$24$\% (15/63) of the clones in the medium group are copy-pastes, and from the large
group we found that $6$\%  (4/63) of the clones are copy-pastes. 
According to the 2-sample test for equality of proportions with continuity correction these difference in proportions are statistically significant (p-value of $0.001$, less than or equal to an alpha of $\alpha = 0.05$).

%
%
%
%

The number of copy-pastes in the medium group seemed reasonable. Upon further
manual review of the clones we found that the number of lines in a clone are not very
meaningful for characterizing copy-pasted code blocks. Because, functions can
often have large blocks of documentation and very little code, e.g. a 20 line
comment with 2 lines of code in 1 function.

For the reader we describe here 3 of the clones from each of the groups to check for license
conflicts.
Here are 3 examples from the medium group.
\textbf{Clone 1} between \path{Aesthete-0.4.2/aesthete/glypher/gutils.py,23,37} (the file,
start line
number, end line number) and StackOverflow (SO) answer 42161514. The Python
file does not contain a license header; but, the package is licensed as GPLv3.
Therefore, this conflicts with the \texttt{CC-BY-SA 3.0} license of StackOverflow.
The copied code is referred to as a universal XML indent function. However,
no library or module is cited in the StackOverflow post.
\textbf{Clone 2} between \path{Argot-0.6/argot/utils.py,34,46} and StackOverflow answer 3271650.
The Python module is licensed under a custom license that is not compatible with \texttt{CC-BY-SA}.
 Therefore, the two code blocks conflict. The copied code provides
an HTML decoding function. It cites a source URL that is no longer available
in the StackOverflow post.
\textbf{Clone 3} between \path{SQLAlchemy-0.9.7/test/ext/declarative/test_basic.py,179,193}
and StackOverflow answer 8297804. The Python module has an MIT license and therefore
conflicts with the StackOverflow code. The code samples both make mock database objects in
Python and are slightly modified from each other. It is not clear if the two
code blocks are intentional copy-pastes or if the two blocks are using
conventional \emph{example} variable names to demonstrate a point.

Here are 3 examples from the large group.
\textbf{Clone 4} between \texttt{adi.slickstyle\--0.1/\-PasteScript\--1.7.5-\-py2.6.egg/\-paste/\-script/\-util/\-sub\-process\-24.\-py\-,762,813} and StackOverflow answer 12965273. The code block is licensed
under an MIT license that conflicts with the StackOverflow code. The two
code blocks are a copied function namely \texttt{subprocess\-.communicate}.
\textbf{Clone 5} between \path{SQLAlchemy-1.0.13/examples/generic_associations/generic_fk.py,42,73} and StackOverflow answer 17757700. The Python block is licensed under an MIT
license that is incompatible with the StackOverflow code. The StackOverflow post has potentially copied an
example class, that represents a street address, out of the pip module.
\textbf{Clone 6} between \path{Agora-Curator-0.1.2/agora/curator/__init__.py,1,32} and
StackOverflow answer 40364576. The Python block is licensed under Apache-2 which is
incompatible with the StackOverflow code. The two code blocks have the same code, but the
pip module contains many lines of commented code. This is effectively a 6 line
code clone instead of 32 lines.

Thus software relicensing may have occurred in these
copy-pastes.
We construct Table~\ref{tab:dist} to show the distributions of licenses
from our detected clones, of querying the 1000m data set against our
StackOverflow data set with the Apprentices.
\emph{NotQuiteNinka} shows that none of our code blocks from the pip dataset were licensed appear to share the \texttt{CC-BY-SA} StackOverflow license. This could mean that there are
many copy-pastes without proper relicensing. However, the StackOverflow code
may have cited its source in the surrounding text of the posts, which
we did not analyze in our evaluation.

\hassanbox{There are many cases of code clones with incompatible licenses and lacking attribution occurring between Python F/LOSS projects and StackOverflow.}

\subsection{\caseD: copy-paste detection with license compliance between documentation and StackOverflow}

We investigate if license conflicts also occur between code blocks in
non-source code based artifacts, such as StackOverflow posts and
Python 2.7 documentation. We compare Python 2.7 documentation with StackOverflow.
We follow a similar methodology to \emph{\caseC}.

We sample 63 of the detected clone pairs randomly to search for
copy-pasted content as in the other case studies. Manually, we go read
through each of the clone pairs and check if the code blocks are
lexically similar and if the clone is type-1, type-2, or type-3. If it
is type-1 then it is a direct copy paste; but, if it is type-2 or
type-3 it is considered copy paste with minor modification.

Afterwards we checked for textual attribution of code clone snippets to
the Python Software Foundation---as per the license of Python 2.7
documentation.

\textbf{The result:} we find that 74.6\% (47/63 sampled clones) of code clones between,
the two sources of code blocks, contain copy-pasted code with minor modification. These code clones have conflicting licenses because the Python
documentation is licensed under the Python software foundation license that is
incompatible with the StackOverflow code blocks' \texttt{CC-BY-SA 3.0} license.
This shows that StackOverflow posters are not properly attributing and licensing the source code in their posts---corrobating Baltes et al.~\cite{baltes2018usage} evidence of lack of license awareness on StackOverflow.

We extracted 5,050,641 Python code blocks from the StackOverflow data
set we collected with our parser. From the Python 2.7's documentation files we
extracted 409,260 code blocks with our parser. We found 68,491 code clones between
Python 2.7 documentation and StackOverflow, with an 80\% similarity
and at least 23 tokens in each code block of the pairs using the Apprentices. This is
interesting because it demonstrates that many code clones come from documentation, not just project source code.

Of the 63 samples that we investigated, we found 23 of these code pairs to be copy pastes and 24 additional code pairs to be type 2 or 3 code clone copy pastes.
Our manual inspection of sampled code clones shows that approximately 74.6\% of these code clones are copy pasted code
with minor modification and it is unclear how these code snippets migrated
between the documentation and StackOverflow.

We also attempted to check the (697,401) Python related StackOverflow posts for
attribution to the Python Software Foundation (PSF) in the case that developers
attempted to attribute reference material they had accessed. We found $2$ user
talked about the PSF in their post not related to a code attribution; $1$ user
copies PSF licensed code into their StackOverflow post without attribution in
question $40972386$;
$1$ user asked how and when code should use the PSF license in
question $16335342$ but this was closed as off-topic;
and $2$ users post PSF licensed code to StackOverflow with incomplete
attributions to the sources in questions $18816421$ and $25638502$.
We can see that 0\% (6/697401) of python posts mentioned the PSF but none of
the observed posts attributed it.
The users show that attributing licensed work is not well understood on
StackOverflow---thus even if licenses were compatible the lack of attribution is a violation of the Python 2.7's documentation license. 

Thus we find that the copying of code between Python documentation and 
StackOverflow is a grave \emph{concern} because StackOverflow's \texttt{CC-BY-SA 3.0} license conflicts with the
Python Software Foundation license applied to the documentation. That means
the unclear code migration between the documentation and StackOverflow likely
contains license violations from StackOverflow users posting documentation to
the web service.
This is exacerbated by the related work which shows that 66\% of StackOverflow users are unaware of the
software license that is applied to the code they post~\cite{baltes2018usage}.
This motivates further 
investigation to check if code migrating to StackOverflow is attributed properly
and is relicensed properly.

\hassanbox{While Yang et al.~\cite{yang2017stack} confirm that code
  flows from StackOverflow into F/LOSS, our results confirm that F/LOSS code
  from projects such as Python flow into StackOverflow where, they are
  relicensed and thus potential license violations.}

\subsection{Evaluation Summary}

We showed that there is copy-pasted code between open source projects and
StackOverflow, which is surprising given the relicensing that occurs
when code is posted on StackOverflow. We also showed that our web service
is capable of detecting copied homework between students and from online
resources. Furthermore, we showed improvements via performance measurements
on the underlying clone detection method used by our web service.

\begin{table}
  \centering
  \begin{tabular}{| l | r | r |}
    \hline
    License & Clone Pairs & Percent of Clones \\ \hline
    \hline
    Total clones & 80,796 & - \\ \hline
    GPLv3+ conflicts & 4,136 & 5.12\%\\ \hline
    Apache-2 conflicts & 2,650 & 3.28\%\\ \hline
    MIT conflicts & 46,709 & 57.81\%\\ \hline  
    Lack of licensing & 11,799 & 14.60\%\\ \hline 
    Unknown license & 9,196 & 11.38\% \\ \hline 
  \end{tabular}
  \caption{A subset of license distribution conflicts in the 1000m queried against StackOverflow data sets. Licenses are from the pip modules. The licenses conflict unless they are \texttt{CC-BY-SA} like the StackOverflow code.
    }
  \label{tab:dist}
\end{table}

\balance

\section{Threats to Validity}

\textbf{Construct Validity:} Construct validity was threatened by
assumptions regarding the licenses. We assumed that licenses can be
applied recursively with NotQuiteNinka, which violates Ninka's very
conservative rate of detection. 

\textbf{Internal Validity:} internal validity was threatened by
provenance. All of the clones evaluated discussed and evaluated were
of ambiguous origin and we did not do the provenance analysis to
determine which is the true. 
Parameters such as 
 minimum clone size could threaten reported measurements.
Internal validity was also biased by our
selection of exercises, as we did not sample, we relied on GitHub
ranks. NotQuiteNinka was not evaluated in this paper for license accuracy.

\textbf{External Validity:} threats to generalizability of our
approach is that we only used python language source code, pip
modules, and Python documentation rather than other programs. Our
programming exercises were not necessarily representative of actual
student assignments. We rely on SourcererCC's definition of a clone
which might differ from other clone detectors.

\section{Conclusion}
\label{conclusion}

In this paper we described the results found by the Sourcerer's Apprentice:
that code is copy and pasted onto StackOverflow without proper relicensing;
that we can use the tool to check for copied homework solution; and we
demonstrate and discuss performance improvements made to underlying
code clone detection techniques.

The Sourcerer's apprentice combines existing license inference tools
such as Ninka with an existing state of the art code clone detector
SourcererCC in a webservice to allow for repeated
querying of large corpus of source code.
We described the engineering efforts in terms of profiling, asynchronous task parallelization, and distributed data sharding that were employed to make a horizontally scalable web-service with reasonable run-time perform.
By distributing apprentices across multiple machines we can improve runtime performance of code clone tasks via parallelization.
By amortizing the heavy start-up time of
loading and indexing we improved query performance. By profiling and
debugging the indexer we improved index performance. By wrapping Ninka
with a recursive license inferencer we inferred licenses for more
files.

We built the Sourcerer's Apprentice because we wanted to gain a better understanding of
code clone migration and the licensing these resulting code blocks have.
How does the license change as code is copied through different medias
and projects, one might ask.
Thus we designed the Sourcerer's
Apprentice to address 4 specific use cases: searching for clones
between StackOverflow posts and programming exercises; searching
for clones between documentation and StackOverflow posts;
searching for
clones within a set of a programming exercises; and searching for
clones between a software project and a larger corpus of projects. We
demonstrated the effectiveness of the tool on each use case.

Through empirical study enabled by Sourcerer's Apprentice we were able to
highlight that the source of much StackOverflow Python source code was not
from StackOverflow itself, but from other software projects. In
particular there were many clones ($68,491$) of the Python
project's documentation source code which did not share the same
license as StackOverflow. Similarly, we found $46,709$ clones from the
1000 pip module data set with an \texttt{MIT} license that conflicts with the
StackOverflow license.
This shows that the flow of code is not
just from StackOverflow to Open Source Software projects on GitHub, but it is
also from Open Source Software projects to StackOverflow. We can also see the
prior work's result that 66\% of developers do not know what license is applied to the
code they post on StackOverflow~\cite{baltes2018usage} to see that developer's need
help tracking the licenses of their code snippets when developing software.

Beyond incompatible licenses, we show nearly \emph{complete lack of attribution} to the Python Software Foundation in Python documentation clones. This means that of the 68,491 Python 2.7 documentation only 4 potentially provide attribution. Therfore, StackOverflow and their users are clearly violating Open Source attribution requirements thus invalidating the use and distribution of the source code they have posted.

\subsection{Recommendations to stakeholders}

\emph{For developers and companies,} we recommend that one checks thoroughly the true provenance of code found on StackOverflow. Not only do StackOverflow posters need attribution via CC-BY-SA 3.0, but the original authors who authored the shared F/LOSS code also require attribution.

\emph{For StackOverflow}, we recommend that
StackOverflow provide tool support to allow for better attribution meta-data, addressing the
reality that not all snippets on StackOverflow can be relicensed, and that StackOverflow is currently distributing thousands of snippets in violation of both license and attribution requirements. It would be in StackOverflow's interest to run a service like the Apprentice to automatically check for mis-attribution or license violations from documentation, and other sources, before they are committed within questions and answers.

\emph{For the Python Software Foundation and other F/LOSS authors}, we
recommend that they pay attention to this problem and potentially make
edits that attribute their foundation on existing StackOverflow
question.

\bibliographystyle{IEEEtran}
\bibliography{references}

\begin{thebibliography}{10}
\providecommand{\url}[1]{#1}
\csname url@samestyle\endcsname
\providecommand{\newblock}{\relax}
\providecommand{\bibinfo}[2]{#2}
\providecommand{\BIBentrySTDinterwordspacing}{\spaceskip=0pt\relax}
\providecommand{\BIBentryALTinterwordstretchfactor}{4}
\providecommand{\BIBentryALTinterwordspacing}{\spaceskip=\fontdimen2\font plus
\BIBentryALTinterwordstretchfactor\fontdimen3\font minus
  \fontdimen4\font\relax}
\providecommand{\BIBforeignlanguage}[2]{{%
\expandafter\ifx\csname l@#1\endcsname\relax
\typeout{** WARNING: IEEEtran.bst: No hyphenation pattern has been}%
\typeout{** loaded for the language `#1'. Using the pattern for}%
\typeout{** the default language instead.}%
\else
\language=\csname l@#1\endcsname
\fi
#2}}
\providecommand{\BIBdecl}{\relax}
\BIBdecl

\bibitem{SO}
S.~E. Inc., ``Stack overflow,'' \url{https://stackoverflow.com/}, 2018,
  [Online; accessed 8-Janurary-2018].

\bibitem{github}
GitHub, ``Built for developers,'' \url{https://github.com/}, 2018, [Online;
  accessed 8-Janurary-2018].

\bibitem{bltj}
J.~Reddy, ``{The Consequences of Violating Open Source Licenses},''
  \url{http://btlj.org/2015/11/consequences-violating-open-source-licenses/},
  2015-11-08, [Online; accessed 12-June-2018].

\bibitem{itworld}
B.~Proffitt, ``{How scary are GPL violations?}''
  \url{https://www.itworld.com/article/2734669/it-management/how-scary-are-gpl-violations-.html},
  2011-11-21, [Online; accessed 12-June-2018].

\bibitem{scompliance}
C.~Maracke, ``{Understanding License Compatibility and Compliance Risks \&
  Processes in Free and Open Source Software},''
  \url{https://static1.squarespace.com/static/51df34b1e4b08840dcfd2841/t/5825fd632e69cfa706674318/1478884745610/ROS-I-Conf2016-day2-08-maracke.pdf},
  2016, [Online; accessed 12-June-2018].

\bibitem{lexology}
B.~.~P. LLP, ``{Copyright infringement of open source software in Canada},''
  \url{https://www.lexology.com/library/detail.aspx?g=9b3e28b0-9412-4f9f-a332-353bead1b840},
  2008-11-04, [Online; accessed 12-June-2018].

\bibitem{yang2017stack}
D.~Yang, P.~Martins, V.~Saini, and C.~Lopes, ``Stack overflow in github: any
  snippets there?'' in \emph{Proceedings of the 14th International Conference
  on Mining Software Repositories}.\hskip 1em plus 0.5em minus 0.4em\relax IEEE
  Press, 2017, pp. 280--290.

\bibitem{SOisCCBYSA3}
angussidney, ``{TL;DR: Source code on SO is still licensed under CC-BY-SA.}''
  \url{https://meta.stackexchange.com/questions/285711/source-code-on-stack-overflow-cc-by-sa-or-mit/285723\#285723},
  2016, [Online; accessed 8-Janurary-2018].

\bibitem{ccbysa3}
C.~Commons, ``{Attribution-ShareAlike 3.0 Unported (CC BY-SA 3.0)},''
  \url{https://creativecommons.org/licenses/by-sa/3.0/}, 2007, [Online;
  accessed 8-Janurary-2018].

\bibitem{baltes2018usage}
S.~Baltes and S.~Diehl, ``Usage and attribution of stack overflow code snippets
  in github projects,'' \emph{arXiv preprint arXiv:1802.02938}, 2018.

\bibitem{sajnani2016sourcerercc}
H.~Sajnani, V.~Saini, J.~Svajlenko, C.~K. Roy, and C.~V. Lopes, ``Sourcerercc:
  Scaling code clone detection to big-code,'' in \emph{Software Engineering
  (ICSE), 2016 IEEE/ACM 38th International Conference on}.\hskip 1em plus 0.5em
  minus 0.4em\relax IEEE, 2016, pp. 1157--1168.

\bibitem{german2010sentence}
D.~M. German, Y.~Manabe, and K.~Inoue, ``A sentence-matching method for
  automatic license identification of source code files,'' in \emph{Proceedings
  of the IEEE/ACM international conference on Automated software engineering},
  ser. ASE '10.\hskip 1em plus 0.5em minus 0.4em\relax New York, NY, USA: ACM,
  2010.

\bibitem{vendome2017license}
C.~Vendome, G.~Bavota, M.~Di~Penta, M.~Linares-V{\'a}squez, D.~German, and
  D.~Poshyvanyk, ``License usage and changes: a large-scale study on github,''
  \emph{Empirical Software Engineering}, pp. 1--41, 2017.

\bibitem{gpllicenses}
F.~S. Foundation, ``Various licenses and comments about them,''
  \url{https://www.gnu.org/licenses/license-list.en.html}, 2018, [Online;
  accessed 8-Janurary-2018].

\bibitem{german2012method}
D.~German and M.~Di~Penta, ``A method for open source license compliance of
  java applications,'' \emph{IEEE software}, vol.~29, no.~3, pp. 58--63, 2012.

\bibitem{german2009license}
D.~M. German and A.~E. Hassan, ``License integration patterns: Addressing
  license mismatches in component-based development,'' in \emph{Proceedings of
  the 31st International Conference on Software Engineering}.\hskip 1em plus
  0.5em minus 0.4em\relax IEEE Computer Society, 2009, pp. 188--198.

\bibitem{van2014tracing}
S.~Van Der~Burg, E.~Dolstra, S.~McIntosh, J.~Davies, D.~M. German, and
  A.~Hemel, ``Tracing software build processes to uncover license compliance
  inconsistencies,'' in \emph{Proceedings of the 29th ACM/IEEE international
  conference on Automated software engineering}.\hskip 1em plus 0.5em minus
  0.4em\relax ACM, 2014, pp. 731--742.

\bibitem{mlouki2016detection}
O.~Mlouki, F.~Khomh, and G.~Antoniol, ``On the detection of licenses violations
  in the android ecosystem,'' in \emph{Software Analysis, Evolution, and
  Reengineering (SANER), 2016 IEEE 23rd International Conference on},
  vol.~1.\hskip 1em plus 0.5em minus 0.4em\relax IEEE, 2016, pp. 382--392.

\bibitem{vendome2017machine}
C.~Vendome, M.~Linares-V{\'a}squez, G.~Bavota, M.~Di~Penta, D.~German, and
  D.~Poshyvanyk, ``Machine learning-based detection of open source license
  exceptions,'' in \emph{Software Engineering (ICSE), 2017 IEEE/ACM 39th
  International Conference on}.\hskip 1em plus 0.5em minus 0.4em\relax IEEE,
  2017, pp. 118--129.

\bibitem{almeida2017software}
D.~A. Almeida, G.~C. Murphy, G.~Wilson, and M.~Hoye, ``Do software developers
  understand open source licenses?'' in \emph{Program Comprehension (ICPC),
  2017 IEEE/ACM 25th International Conference on}.\hskip 1em plus 0.5em minus
  0.4em\relax IEEE, 2017, pp. 1--11.

\bibitem{roy2007survey}
C.~K. Roy and J.~R. Cordy, ``A survey on software clone detection research,''
  \emph{Queen’s School of Computing TR}, vol. 541, no. 115, pp. 64--68, 2007.

\bibitem{kapser2006cloning}
C.~Kapser and M.~W. Godfrey, ````cloning considered harmful'' considered
  harmful,'' in \emph{Reverse Engineering, 2006. WCRE'06. 13th Working
  Conference on}.\hskip 1em plus 0.5em minus 0.4em\relax IEEE, 2006, pp.
  19--28.

\bibitem{kapser2008cloning}
C.~J. Kapser and M.~W. Godfrey, ````cloning considered harmful'' considered
  harmful: patterns of cloning in software,'' \emph{Empirical Software
  Engineering}, vol.~13, no.~6, p. 645, 2008.

\bibitem{german2009code}
D.~M. German, M.~Di~Penta, Y.-G. Gueheneuc, and G.~Antoniol, ``Code siblings:
  Technical and legal implications of copying code between applications,'' in
  \emph{Mining Software Repositories, 2009. MSR'09. 6th IEEE International
  Working Conference on}.\hskip 1em plus 0.5em minus 0.4em\relax IEEE, 2009,
  pp. 81--90.

\bibitem{githubtos}
GitHub, ``Github terms of service,''
  \url{https://help.github.com/articles/github-terms-of-service/}, 2018,
  [Online; accessed 9-Janurary-2018].

\bibitem{debiandisclaimer}
S.~in~the Public~Interest, ``Disclaimer of the debian mailing lists,''
  \url{https://www.debian.org/MailingLists/disclaimer}, 2018, [Online; accessed
  9-Janurary-2018].

\bibitem{REST}
R.~T. Fielding, ``{Architectural Styles and the Design of Network-based
  Software Architectures},''
  \url{http://www.ics.uci.edu/~fielding/pubs/dissertation/rest_arch_style.htm},
  2000, [Online; accessed 12-Janurary-2018].

\bibitem{kamiya2002ccfinder}
T.~Kamiya, S.~Kusumoto, and K.~Inoue, ``Ccfinder: a multilinguistic token-based
  code clone detection system for large scale source code,'' \emph{IEEE
  Transactions on Software Engineering}, vol.~28, no.~7, pp. 654--670, 2002.

\bibitem{jiang2007deckard}
L.~Jiang, G.~Misherghi, Z.~Su, and S.~Glondu, ``Deckard: Scalable and accurate
  tree-based detection of code clones,'' in \emph{Proceedings of the 29th
  international conference on Software Engineering}.\hskip 1em plus 0.5em minus
  0.4em\relax IEEE Computer Society, 2007, pp. 96--105.

\bibitem{gode2009incremental}
N.~G{\"o}de and R.~Koschke, ``Incremental clone detection,'' in \emph{Software
  Maintenance and Reengineering, 2009. CSMR'09. 13th European Conference
  on}.\hskip 1em plus 0.5em minus 0.4em\relax IEEE, 2009, pp. 219--228.

\bibitem{cordy2011nicad}
J.~R. Cordy and C.~K. Roy, ``The nicad clone detector,'' in \emph{Program
  Comprehension (ICPC), 2011 IEEE 19th International Conference on}.\hskip 1em
  plus 0.5em minus 0.4em\relax IEEE, 2011, pp. 219--220.

\bibitem{wang2012can}
X.~Wang, Y.~Dang, L.~Zhang, D.~Zhang, E.~Lan, and H.~Mei, ``Can i clone this
  piece of code here?'' in \emph{Proceedings of the 27th IEEE/ACM International
  Conference on Automated Software Engineering}.\hskip 1em plus 0.5em minus
  0.4em\relax ACM, 2012, pp. 170--179.

\bibitem{lin2015clone}
Y.~Lin, X.~Peng, Z.~Xing, D.~Zheng, and W.~Zhao, ``Clone-based and interactive
  recommendation for modifying pasted code,'' in \emph{Proceedings of the 2015
  10th Joint Meeting on Foundations of Software Engineering}.\hskip 1em plus
  0.5em minus 0.4em\relax ACM, 2015, pp. 520--531.

\bibitem{toomim2004managing}
M.~Toomim, A.~Begel, and S.~L. Graham, ``Managing duplicated code with linked
  editing,'' in \emph{Visual Languages and Human Centric Computing, 2004 IEEE
  Symposium on}.\hskip 1em plus 0.5em minus 0.4em\relax IEEE, 2004, pp.
  173--180.

\bibitem{kim2009discovering}
M.~Kim and D.~Notkin, ``Discovering and representing systematic code changes,''
  in \emph{Proceedings of the 31st International Conference on Software
  Engineering}.\hskip 1em plus 0.5em minus 0.4em\relax IEEE Computer Society,
  2009, pp. 309--319.

\bibitem{duala2007tracking}
E.~Duala-Ekoko and M.~P. Robillard, ``Tracking code clones in evolving
  software,'' in \emph{Proceedings of the 29th international conference on
  Software Engineering}.\hskip 1em plus 0.5em minus 0.4em\relax IEEE Computer
  Society, 2007, pp. 158--167.

\bibitem{lin2014clonepedia}
Y.~Lin, Z.~Xing, X.~Peng, Y.~Liu, J.~Sun, W.~Zhao, and J.~Dong, ``Clonepedia:
  summarizing code clones by common syntactic context for software
  maintenance,'' in \emph{Software Maintenance and Evolution (ICSME), 2014 IEEE
  International Conference on}.\hskip 1em plus 0.5em minus 0.4em\relax IEEE,
  2014, pp. 341--350.

\bibitem{kapser2004aiding}
C.~Kapser and M.~W. Godfrey, ``Aiding comprehension of cloning through
  categorization,'' in \emph{Software Evolution, 2004. Proceedings. 7th
  International Workshop on Principles of}.\hskip 1em plus 0.5em minus
  0.4em\relax IEEE, 2004, pp. 85--94.

\bibitem{inoue2012does}
K.~Inoue, Y.~Sasaki, P.~Xia, and Y.~Manabe, ``Where does this code come from
  and where does it go?-integrated code history tracker for open source
  systems,'' in \emph{Proceedings of the 34th International Conference on
  Software Engineering}.\hskip 1em plus 0.5em minus 0.4em\relax IEEE Press,
  2012, pp. 331--341.

\bibitem{livieri2007very}
S.~Livieri, Y.~Higo, M.~Matushita, and K.~Inoue, ``Very-large scale code clone
  analysis and visualization of open source programs using distributed
  ccfinder: D-ccfinder,'' in \emph{Software Engineering, 2007. ICSE 2007. 29th
  International Conference on}.\hskip 1em plus 0.5em minus 0.4em\relax IEEE,
  2007, pp. 106--115.

\bibitem{wu2015method}
Y.~Wu, Y.~Manabe, T.~Kanda, D.~M. German, and K.~Inoue, ``A method to detect
  license inconsistencies in large-scale open source projects,'' in
  \emph{Proceedings of the 12th Working Conference on Mining Software
  Repositories}.\hskip 1em plus 0.5em minus 0.4em\relax IEEE Press, 2015, pp.
  324--333.

\bibitem{wu2017analysis}
------, ``Analysis of license inconsistency in large collections of open source
  projects,'' \emph{Empirical Software Engineering}, pp. 1--29, 2017.

\bibitem{wu2017developers}
Y.~Wu, Y.~Manabe, D.~M. German, and K.~Inoue, ``How are developers treating
  license inconsistency issues? a case study on license inconsistency evolution
  in foss projects,'' in \emph{IFIP International Conference on Open Source
  Systems}.\hskip 1em plus 0.5em minus 0.4em\relax Springer, 2017, pp. 69--79.

\bibitem{davies2011software}
J.~Davies, D.~M. German, M.~W. Godfrey, and A.~Hindle, ``Software bertillonage:
  finding the provenance of an entity,'' in \emph{Proceedings of the 8th
  working conference on mining software repositories}.\hskip 1em plus 0.5em
  minus 0.4em\relax ACM, 2011.

\bibitem{asttokens}
D.~Sagal, ``Asttokens,'' \url{https://github.com/gristlabs/asttokens}, 2017,
  [Online; accessed 12-Janurary-2018].

\bibitem{pip}
PyPA, ``pip,'' \url{https://pip.pypa.io/en/stable/}, 2016, [Online; accessed
  8-Janurary-2018].

\bibitem{SOData}
S.~E. Inc., ``Stack exchange data dump,''
  \url{https://archive.org/details/stackexchange}, 2014, [Online; accessed
  6-April-2018].

\bibitem{githubPython}
P.~S. Foundation, ``The python programming language,''
  \url{https://github.com/python/cpython/tree/2.7}, 2018, [Online; accessed
  6-April-2018].

\bibitem{pypi}
------, ``Pypi -- the python package index,''
  \url{https://pypi.python.org/simple/}, 2018, [Online; accessed 6-April-2018].

\bibitem{streams}
R.-G. Urma, ``{Processing Data with Java SE 8 Streams},''
  \url{http://www.oracle.com/technetwork/articles/java/ma14-java-se-8-streams-2177646.html},
  2014, [Online; accessed 12-Janurary-2018].

\end{thebibliography}

\end{document}